\documentclass[twocolumn,prb]{revtex4}
\usepackage{graphicx}% Include figure files
\usepackage{dcolumn}% Align table columns on decimal point
\usepackage{bm}% bold math

\begin{document}

\title{Ferromagnetism and interlayer exchange coupling in short period (Ga,Mn)As/GaAs superlattices}

\author{R. Mathieu and P. Svedlindh}
\affiliation{Department of Materials Science, Uppsala University, Box 534, SE-751 21 Uppsala, Sweden}

\author{J. Sadowski}
\affiliation{Niels Bohr Institute, Copenhagen University, DK-2100 Copenhagen, Denmark\linebreak Chalmers University of Technology, SE-412 96 G\"oteborg, Sweden\linebreak Institute of Physics Polish Academy of Sciences, PL-02668 Warszawa, Poland}

\author{K. \'Swiatek}
\affiliation{Institute of Physics Polish Academy of Sciences, PL-02668 Warszawa, Poland}

\author{M. Karlsteen, J. Kanski, and L. Ilver}
\affiliation{Department of Experimental Physics, Chalmers University of Technology and   G\"oteborg University, SE-412 96 G\"oteborg, Sweden}

\date{\today}

\begin{abstract}
Magnetic properties of (Ga,Mn)As/GaAs superlattices are investigated. The structures contain magnetic (Ga,Mn)As layers, separated by thin layers of non-magnetic GaAs spacer. The short period Ga$_{0.93}$Mn$_{0.07}$As/GaAs superlattices exhibit a paramagnetic-to-ferromagnetic phase transition close to 60K, for thicknesses of (Ga,Mn)As down to 23 \AA. For Ga$_{0.96}$Mn$_{0.04}$As/GaAs superlattices of similar dimensions, the Curie temperature associated with the ferromagnetic transition is found to oscillate with the thickness of non magnetic spacer. The observed oscillations are related to an interlayer exchange interaction mediated by the polarized holes of the (Ga,Mn)As layers.
\end{abstract}

\maketitle

\section{Introduction}

The recent discovery of carrier mediated ferromagnetism diluted magnetic semiconductors (DMS) such as (In,Mn)As\cite{ref1,ref2} and (Ga,Mn)As\cite{ref6,ref7} has due to potential applications in magnetic memories, spin injection and quantum computing devices generated intense interest. In this context low dimensional structures are particularly interesting and superlattices (SL) of thin ferromagnetic semiconductors layers separated by non-magnetic spacers are presently in focus. In previous studies of superlattices with (Ga,Mn)As as the magnetic layer and (Al,Ga)As or (In,Ga)As as non-magnetic spacers\cite{refold13,refold15}, the ferromagnetic properties were lost decreasing the thickness of the magnetic layer below 50 \AA. Also, a recent Monte Carlo simulation study of the magnetic order resulting from the indirect exchange interaction between magnetic moment in a AlAs/(Ga,Mn)As quantum well, obtained a lower limit for the width of the quantum well below which no ferromagnetism was observed.\cite{montecarlo1,montecarlo2}.\\ 
Recently, ferromagnetic ordering was detected in AlAs/(Ga,Mn)As heterostructures\cite{newref} containing (Ga,Mn)As layers thinner than 50 \AA. In the present investigation, we observe 1) a paramagnetic-to-ferromagnetic phase transition for (Ga,Mn)As/GaAs superlattices with a thickness of (Ga,Mn)As as small as 23 \AA [i.e. 8 molecular layers (ML)] and  2) an oscillatory interlayer exchange interaction in Ga$_{0.96}$Mn$_{0.04}$As/GaAs superlattices with varying thickness of GaAs spacer.

\section{Experimental}

Two sets of superlattices of the type (Ga,Mn)As$_{(nML)}$/GaAs$_{(mML)}$ were produced by Molecular beam epitaxy (MBE). Set A is composed of SL with (Ga,Mn)As layers containing 7 \% of Mn, and n=8,10,12ML/m=4ML (denoted as 8/4, 10/4 and 12/4), as well a structure with n=12ML/m=6ML (12/6) for comparison. The SL were grown at $T_g$=200$^{o}$C. Set B is constituted of SL with (Ga,Mn)As layers containing 4 \% of Mn, a fixed m=8ML, and varying n. SL with n=1,3,5,7 and 9ML were prepared at $T_g$=200$^{o}$C, while SL with n=4,6,8 and 10 were grown at $T_g$=220$^{o}$C.\\
The SL were grown in an MBE system (KRYOVAK) equipped with an As$_2$ valved cracker source. As substrates we used semi-insulating, epiready GaAs(100) AXT wafers. The MBE growth of superlattice structures was preceded by standard high-temperature (HT) growth of a 5000 \AA thick GaAs buffer. Then the substrate temperature was lowered to $T_g$ and stabilized for 1 hour. Afterward SL structures were grown with 30 s growth interruptions at each interface. For all the samples the number of repetitions was equal to 100. The quality of the SL was checked by X-ray diffraction (XRD). The recorded (004) Bragg reflections were well reproduced by calculations\cite{mbeconf}, and the parameters of each SL structure (thickness of the constituent layers and (Ga,Mn)As composition) deduced from the fittings are in good accordance with the ones measured by RHEED oscillations during the film growth\cite{mbeconf}.\\
Magnetization measurements were performed using a Quantum Design MPMS5 Superconducting QUantum Interference Device (SQUID). The zero field cooled (ZFC) and field cooled (FC) magnetizations of the SL were recorded in a small magnetic field ($H$=20Oe), applied in the plane of the superlattices.

\section{Results and discussion}

\begin{figure}
\includegraphics[scale=0.50]{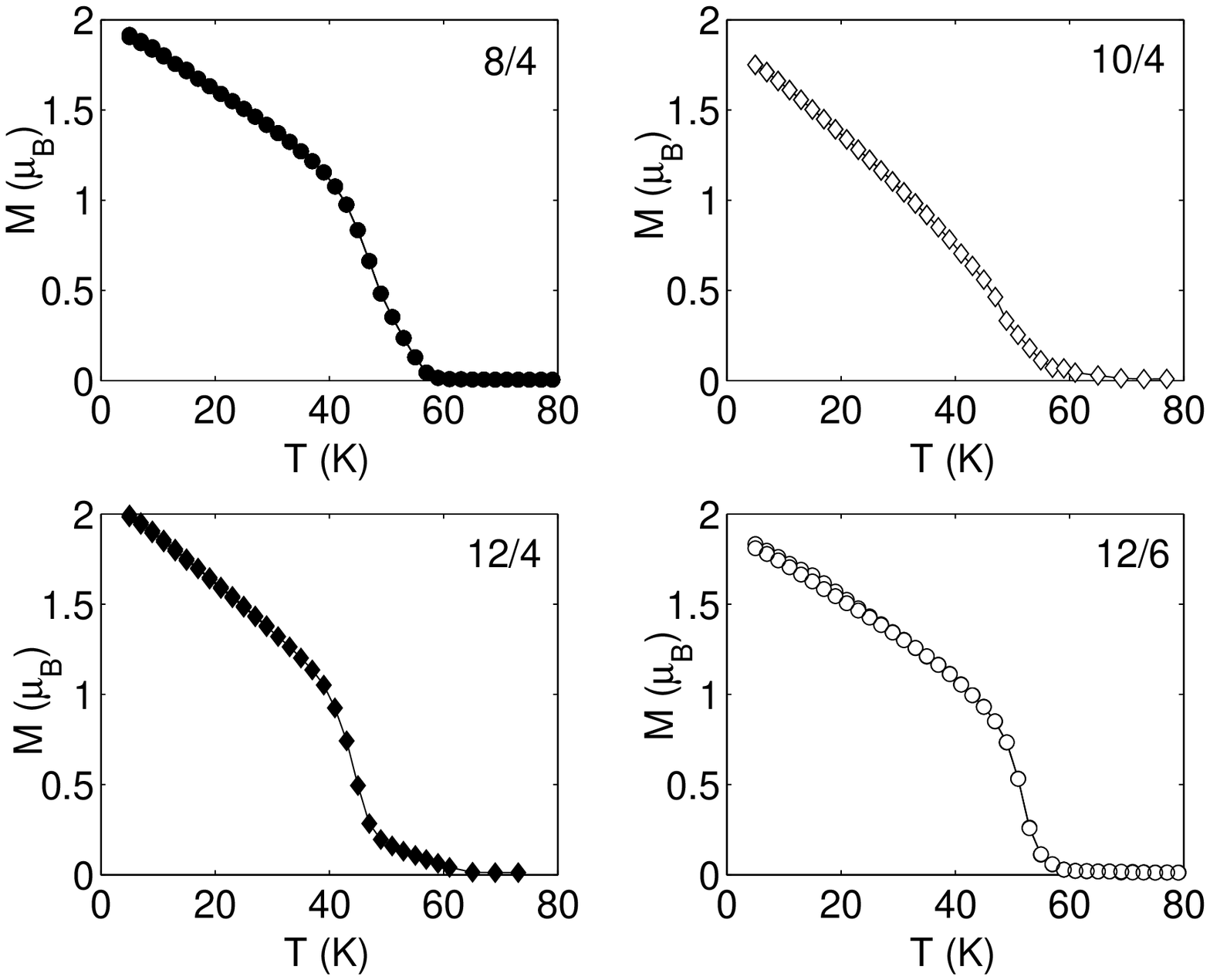}
\caption{Temperature dependence of the ZFC and FC magnetization for set A (superlattices with 7 \% of Mn); $H$=20 Oe.}
\label{fig1}
\end{figure}

\begin{figure}
\includegraphics[scale=0.50]{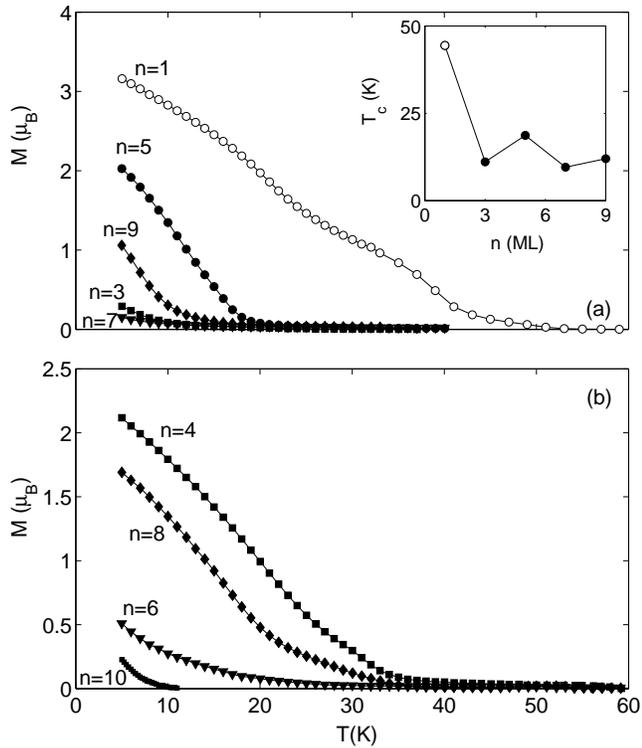}
\caption{Temperature dependence of the FC magnetization for set B (superlattices with 4 \% of Mn, grown at (a) 200$^{o}$C and (b) 220$^{o}$C); $H$=20 Oe. The inset depicts the oscillation of $T_C$ with the number of GaAs monolayers for the SL grown at 200$^{o}$C.} 
\label{fig2}
\end{figure}

Figure\ref{fig1} shows the temperature dependence of the ZFC and FC magnetizations for the first set of superlattices (set A). The non-Brillouin like appearance of the $M(T)$ curves is typical of ferromagnetic (Ga,Mn)As samples, and in this respect, there is little difference comparing the results obtained for the present SL and those obtained for thick (Ga,Mn)As reference samples\cite{refAPL}. Moreover, the ZFC and FC curves coincide in almost the entire  temperature range investigated, indicating low coercivity high quality films. The SL exhibit clear paramagnetic-to-ferromagnetic transitions around $T_C$ $\approx$ 60K (In these systems, it is common to define $T_c$ as the temperature were some onset of magnetic ordering appears). The ferromagnetic ordering of SL with similar dimensions has also been observed using neutron diffraction\cite{ND}, as well as FMR measurements\cite{refPASP}. Transport measurements on this set of SL reveal a metal-insulator transition occurring close to $T_C$\cite{roland}, as observed in ferromagnetic single layers\cite{roland,refAPL}, as well as a relatively large magnetoresistance. In addition, the 12/6 SL exhibits almost identical characteristics comparing to the 8/4 SL; it can be seen on Fig.~\ref{fig1} that 8/4 et 12/6, with the same average Mn content of 4.67 \% have very similar magnetization vs. temperature curves. This indicates that for this serie of SL, a ``digital alloy'' model\cite{kawakami} applies, implying that two SL having the same average Mn concentration in the structure will exhibit similar magnetic properties. The same applies when comparing 8/4 to 10/4 and 12/4: The temperature onset of ferromagnetic behavior increases from 8/4 to 12/4, following the average Mn concentration, which increases from 8/4 to 12/4, as observed for single layers of (Ga,Mn)As\cite{ref6,refAPL,refPASP}. The ``digital alloy'' model for the second set of superlattices (set B): As seen in Fig.\ref{fig2}, the SL with 4 \% Mn and 8/4 (c.f. Fig.~\ref{fig2} (b)) and thus an average  Mn content of about 2.5 \% has a $T_c$ around 35K, close to the value obtained for single layers\cite{refAPL}.\\

Additional effects related to the superlattice structure are evidenced. The paramagnetic-to-ferromagnetic transition of very thin single layers of (Ga,Mn)As deposited on a GaAs substrate is not as sharp as that of thick single layers. The $M$($T$) curves of such thin layers\cite{jannew} (with thicknesses below 100 \AA) are instead very broad\cite{jannew2}. This broadness could be related to some degree of magnetic inhomogeneity, as observed in (Ga,Mn)As layers containing very large amount of Mn\cite{refAPL}. Such inhomogeneity implies that some regions of the sample have stronger coupling than others, yielding the observed broad magnetization curves, as predicted theoretically by Berciu and Bhatt\cite{Berciu}.
In the case of the present superlattices, containing even thinner layers of (Ga,Mn)As, the $M$($T$) curves appear sharper. These curves are, as mentioned above, similar to those obtained for thick (Ga,Mn)As single layers. Boselli et al.\cite{montecarlo2} showed in Monte Carlo simulations that in such a superlattice structure, the holes with spin down (compared to the magnetic layer) were confined within the layer, so that enough holes were available to establish a ferromagnetic interaction in the system; the obtained hole concentration increasing with the number of repetitions in the SL. A coupling between different (Ga,Mn)As layers could also strengthen the ferromagnetic interaction in the structure. Looking again at Fig.~\ref{fig2}, one observes that the ferromagnetic-to-paramagnetic transition temperature of the SL of set B oscillates with the thickness of the non-magnetic spacer layer. These oscillations are similar to those observed in metallic multilayers\cite{metal}. In these systems, magnetic layers are coupled through a nonmagnetic metal, and the interlayer exchange coupling depends on the thickness of the nonmagnetic spacer\cite{bruno}. The oscillations of this coupling, and thus of $T_C$, are usually related to Friedel oscillations created by magnetic impurity atoms in non-magnetic hosts. The Ruderman-Kittel-Kasuya-Yosida (RKKY) model predicts that the polarized charge carriers will mediate a magnetic interlayer interaction varying with the distance $r$ as $J(r) \propto (1/r^2) sin(kr)$, where $k$ is a constant. The oscillations of $T_C$ in the present sets of superlattices could thus be related to a RKKY-like\cite{jung} oscillating interlayer interaction mediated by polarized holes through the GaAs layers.

\section{conclusion}

The magnetic properties of two sets of short period (Ga,Mn)As/GaAs superlattices have been investigated. Contrary to earlier reports, a ferromagnetic interaction is observed in superlattices with (Ga,Mn)As thicknesses down to 23 \AA. In addition, signatures of an interlayer exchange coupling between the magnetic layers of the structure are observed. 

\begin{acknowledgments}

Financial support from the Swedish natural Science Research Council (NFR).

\end{acknowledgments}

\end{document}